\documentclass[12pt]{iopart}

\usepackage{graphicx}
\usepackage{dcolumn}
\usepackage{bm}
\usepackage{multirow}
\usepackage{iopams}
\usepackage{soul}

\begin{document}


\title{Triadic closure dynamics drives scaling-laws in social multiplex networks}

\author{Peter Klimek$^1$, Stefan Thurner$^{1,2,3}$}
\ead{stefan.thurner@meduniwien.ac.at}
\address{
$^1$Section for Science of Complex Systems; Medical University of Vienna; Spitalgasse 23; A-1090; Austria. 
$^2$Santa Fe Institute; 1399 Hyde Park Road; Santa Fe; NM 87501; USA. 
$^3$IIASA, Schlossplatz 1, A-2361 Laxenburg; Austria.
}%

\begin{abstract}
Social networks exhibit scaling-laws for several structural characteristics, such as the degree distribution, the scaling of the attachment kernel, 
and the clustering coefficients as a function of node degree.
A detailed understanding if and how these scaling laws are inter-related is missing so far, 
let alone whether they can be understood through a common, dynamical principle. 
We propose a simple model for stationary network formation and show that the three mentioned scaling relations
follow as natural consequences of triadic closure.
The validity of the model is tested on multiplex data from a well studied massive multiplayer online game.
We find that the three scaling exponents observed in the multiplex data for the  friendship, communication and trading networks 
can simultaneously be explained by the model.
These results suggest that triadic closure could be identified as one of the fundamental dynamical principles 
in social multiplex network formation.
\end{abstract}

\maketitle

Social networks often exhibit statistical structures that manifest themselves in scaling-laws which 
can be quantified through a set of characteristic exponents. 
Maybe the three most relevant scaling laws in terms of network formation are the linking probability for 
new nodes joining the network as a function of degree of the existing (linked-to) node, 
the degree distribution, and the clustering coefficient of nodes as a function of their degree. 
In particular, the probability for a node to acquire a new link, the {\em attachment kernel} $\Pi(k)$, 
often scales with the node degree $k$ \cite{Newman01a, Jeong03} as
\begin{equation}
\Pi(k) \propto k^{\gamma} \quad.
\label{ScalingPi}
\end{equation}
The degree distribution of social networks, i.e. the probability to find a node with a given degree $k$, $P(k)$, often shows 
features of exponential, fat-tailed distributions \cite{Barabasi99, Newman01} or something inbetween, 
depending on the type of social interaction \cite{Onnela07, Szell10}. 
They can be parameterized conveniently by the $q$-exponential \cite{Thurner05,Thurner07},  
\begin{equation}
P(k) \propto \left( 1 + (1-q) k \right)^{\frac{1}{1-q}} \quad,
\label{ScalingK}
\end{equation}
with $q$ a parameter that determines an asymptotic scaling exponent $1/(1-q)$.
A third scaling law, which is ubiquitous in social networks \cite{Onnela07, Szell10, Vazquez02, Ravasz03}, 
is observed for the clustering coefficients $c(k)$ as function of node degree,
\begin{equation}
c(k) \propto k^{-\beta} \quad.
\label{ScalingC}
\end{equation}

Despite the overwhelming empirical evidence for the scaling laws in equations (\ref{ScalingPi} - \ref{ScalingC}), 
it is still undecided if they share a common dynamical origin, and if and how characteristic exponents are related to each other.
For example, for growing network models, where new nodes are constantly added which link through a preferential attachment rule 
to already existing nodes \cite{Barabasi99}, a relation between scaling exponents of the degree distribution and the 
attachment kernel $\gamma$ has been found \cite{Krapivsky00}.
However, these models can not explain the observed scaling of the clustering coefficients.
Moreover, the preferential attachment process \cite{Barabasi99} requires {\it global} information 
(the degrees of all nodes in the network) to establish a new social tie, which is clearly an unrealistic 
assumption for most social networks.
To overcome this problem, growth and preferential attachment mechanisms have been 
extended by {\it local} network formation rules \cite{Jin01, Holme02, Vazquez03, Li10}, where 
a node's linking dynamics only depends on its neighbors or second neighbors.
One such local rule which is extremely relevant for social network formation is the 
principle of {\em triadic closure} \cite{Rapoport53, Granovetter73}, which means that the probability of 
a new link to close a triad is higher than the probability to connect any two nodes.
Scaling-laws for the degree distribution \cite{Holme02}, degree distribution and clustering coefficients \cite{Vazquez03, Toivonen06}, and preferential attachment \cite{Li10} have been reproduced in the 
context of specific models using triadic closure, respectively.
While it is instructive to see how a combination of growth, preferential 
attachment and clustering processes give rise to the three scaling laws above,
this does not help us to understand if the existence and possible inter-relations of the three exponents can emerge from a 
single underlying dynamical origin, and to which extent this common origin is an actual feature of real social network formation processes.
Less is known on relations between characteristic exponents in non-growing, stationary networks \cite{Thurner05, Kim05}.
It has been shown that triadic closure is related to scaling-laws for the degree distribution and  clustering coefficients 
in the stationary case \cite{Davidsen02, Marsili04, Kumpula07, Toivonen09}.

Here we study a simple model that simultaneously explains the three scaling laws in equations (\ref{ScalingPi} - \ref{ScalingC}) 
based on the process of triadic closure in non-growing networks.
This process introduces a mechanism from which preferential attachment emerges, leads to fat-tailed degree distributions, 
and induces scaling of the clustering coefficients with node degrees.
The model is validated with data from a social multiplex, i.e. a superposition of several social networks 
labeled by $\alpha$ with adjacency matrices $M_{\alpha}$, defined on the same set of nodes \cite{Wassermann94}.
The model can be fully calibrated with the multiplex data and explains three observed characteristic 
exponents for three different sub-networks of the multiplex.

\begin{figure}[tbp]
 \begin{center}
 \includegraphics[width=65mm]{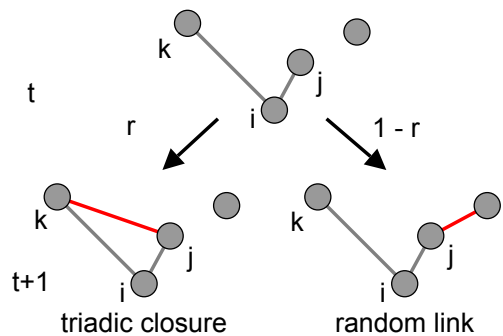}
  \end{center}
 \caption{Node $i$ (with more than two links) and one of its neighbors $j$ are randomly selected. 
 With probability $r$ the process of triadic closure takes place (triad consists of $i,j,k$), with probability $1-r$, $j$ links to a random node.}
  \label{scheme}
 \end{figure}

\section{\label{sec:dandm}Results}

\subsection{Model specification}

The model is built around the process of triadic closure, the principle that links tend to be created between 
nodes that share a neighbor. The model  includes the addition and removal of nodes.
The network is initialized with $N$ nodes, each node having one link to a randomly chosen node.
The dynamics is completely specified by an iteration of the following steps, starting at $t$.
\begin{enumerate}
\item Pick a node $i$ at random. If $i$ has less than two links, create a link between $i$ 
and any randomly chosen node, and continue with step (iii). 
If $i$ has two or more links, choose one of its neighbors at random, say node $j$, and continue with step (ii).
\item With probability $r$ (triadic closure parameter), create a link between $j$ and another randomly 
chosen neighbor of $i$, say $k$. With probability $1-r$, create a link between $j$ and a node  randomly chosen 
from the entire network, see figure \ref{scheme}.
\item With probability $p$ (node-turnover parameter) remove a randomly chosen node from the network along with all its links, and introduce a new node linking to $m$ randomly chosen nodes. Then continue with time-step $t+1$.
\end{enumerate}
For $p>0$ nodes have a finite lifetime, which implies that the network reaches a stationary state 
where the total number of links $L(t)$ and the network measures  $\Pi(k)$, $P(k)$, and $c(k)$ 
fluctuate around steady state levels. 
The model is a variant of the model proposed in \cite{Davidsen02}, which is contained as the 
special case $r=1$ in the above protocol. Our model can also be seen as a stationary 
version of the connecting-nearest-neighbors-model in \cite{Vazquez03}.
Combinations of triadic closure and random edge attachment have also been studied in growing \cite{Holme02, Toivonen06}, and weighted \cite{Kumpula07} networks.
Reaching a stationary state is independent of $m$. 
The model is completely specified by four parameters, $N$, $r$, $p$, and $m$.   

\begin{figure}[t]
 \begin{center}
 \includegraphics[width=170mm]{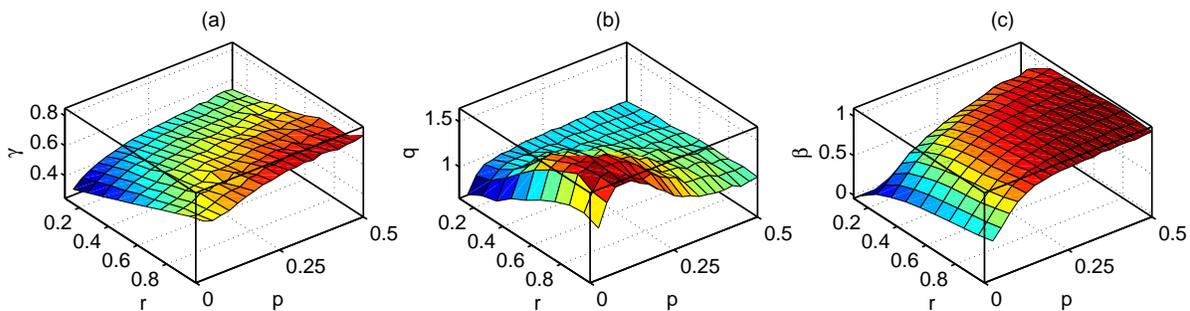}
  \end{center}
 \caption{Dependence of scaling exponents $\gamma$, $q$, and $\beta$ on the model parameters $p$ and $r$. 
 (a) $\gamma$ becomes closer to one for high $p$ or $r$, and is confined to the interval $0<\gamma<1$. 
 (b) $q$ is large for small $p$ and large $r$, and approaches one for large $p$. 
 (c) $\beta$ is close to zero for $r$ close to zero, and approaches $\beta = -1$ for large values of $p$ and $r$.}
  \label{slopes}
 \end{figure}

\begin{table}[htb]
\caption{Summary of network measures and model results. For the Pardus friendship ($\alpha=1$), 
communication (comm., $\alpha=2$), and trade ($\alpha=3$) networks the number of nodes $N_{\alpha}$, links $L_{\alpha}$, 
average degree $\bar k_{\alpha}$, and  average number of nodes entering and leaving the network per day, 
$\Delta n^+_{\alpha}$ and $\Delta n^-_{\alpha}$, are shown. The results of the calibration of the model to the empirical 
networks, $r$ and $p$, are given, together with the fit results of the parameters $\gamma$, $q$, and $\beta$ for the 
data and the model.}
\setlength{\tabcolsep}{2pt}
\begin{tabular}{lr|rrrrr|rr|rrrrrr}
\multicolumn{2}{c|}{type} & \multicolumn{5}{c|}{network features} & \multicolumn{2}{c|}{parameter} & \multicolumn{6}{c}{exponents (data and model)}
\\ 
  & $\alpha$ & $N_{\alpha}$ & $L_{\alpha}$ & $\bar k_{\alpha}$ & $\Delta n^+_{\alpha}$ & $\Delta n^-_{\alpha}$ & $r_{\alpha}$ & $p_{\alpha}$ & $\gamma$ & $\gamma_{mod}$ & $q$ & $q_{mod}$ & $\beta$ & $\beta_{mod}$ \\ \hline
friends & 1 & 4,547 &  21,622 &  9.5 & 24.26 & 23.07 & 0.58 & 0.12 & 0.88(4) & 0.77(2) & 1.16(1) & 1.116(2) & 0.69(3) & 0.66(3)  \\
comm. & 2 & 2,810 & 9,420 & 6.7 & 110.2 & 109.4 & 0.57 & 0.18 & 0.84(1) & 0.76(2) & 1.24(1) & 1.148(3) & 0.59(3) & 0.78(3) \\
trade & 3 & 4,514 & 31,475 & 13.9 & 58.58 & 56.19 & 0.80 & 0.08 & 0.83(1) & 0.80(1) & 1.073(1) & 1.102(1) & 0.63(3) & 0.60(3)
\end{tabular}
\label{nwtab}
\end{table}
 
\subsection{Estimation of model parameters}

Social ties are often established between two individuals by being introduced by a mutual acquaintance.
Other modes of social tie formation, such as random encounters may not lead to triadic closure.
Step (ii) in the above protocol captures these two linking processes.
Ties also change because people enter and leave social circles, for example they change workplaces, 
move to different cities, or change their hobbies. 
This is incorporated in step (iii). To calibrate the model to a real social multiplex network, 
$M_{\alpha}$ with $N_{\alpha}$ nodes and $L_{\alpha}$ links, the stationarity assumption 
has to be checked, and the parameters for triadic closure $r$, and node-turnover $p$ have to be estimated.
Consider the average number of nodes entering ($\Delta n^+_{\alpha}$) and leaving ($\Delta n^-_{\alpha}$) 
the network $M_{\alpha}$ per time unit.
For  stationarity  to hold we demand 
\begin{equation}
\Delta n^+_{\alpha} \approx \Delta n^-_{\alpha} \gg \Delta n^+_{\alpha} - \Delta n^-_{\alpha} \quad,
\label{stationarity}
\end{equation}
i.e.  the net growth rate is much smaller than the rates at which nodes enter or leave the network.
The triadic closure parameter $r_{\alpha}$ can be directly measured as the ratio between 
the number of {\em links} in network $M_{\alpha}$ which -- at their creation -- close at least one triangle, 
and the total number of created links. 
The node-turnover parameter $p$ can be estimated by demanding for the number of links in the model and in the 
real network to be the same. To see this, note that one adds on average $\Delta l^+$  
and removes $\Delta l^-$ links per time-step. Stationarity means that $\Delta l^+=\Delta l^-$.
Since one link is created at each time-step in either step (i) or (ii), and with probability $p$, 
$m$ links are added in step (iii), we have $\Delta l^+ = 1+pm$.
Denoting the average degree by $\bar k = \frac{2 N}{L}$, with probability $p$, in step (iii), one removes on average $\bar k$ links 
per time-step, $\Delta l^- = p \bar k$.
To calibrate the model to a network $M_{\alpha}$ the turnover parameter $p_{\alpha}$ is
\begin{equation}
p_{\alpha} = \frac{1}{\bar k_{\alpha}-m} \quad.
\label{turnover}
\end{equation}
The model is initialized with $N_{\alpha}$ nodes and the dynamics follows the protocol with parameters $r_{\alpha}$ and $p_{\alpha}$.
After a transient phase the number of links fluctuates around $L_{\alpha}$, and the scaling exponents $\gamma, q, \beta$ approach stationary values.

Calibration of the model requires complete, time-resolved topological information $M_{\alpha}(t)$ 
over a large number of link-creation processes. Suitable data is available for example in the social multiplex 
network of the online game 'Pardus' \cite{Szell10, Szell10SN, SzellChapter, Thurner12, Szell12}, see the Methods section. 
Table \ref{nwtab} summarizes key features of  $M_{\alpha}$, including the number of nodes $N_{\alpha}$,  
links $L_{\alpha}$ for the Pardus friendship ($\alpha=1$), communication ($\alpha=2$), and trade ($\alpha=3$) networks. 
Table \ref{nwtab} also lists the average degree $\bar k_{\alpha}$, as measured on the last day of the observation record, 
and the average number of nodes entering ($\Delta n^+_{\alpha}$) and leaving ($\Delta n^-_{\alpha}$) per day, confirming that the networks are in fact stationary in the sense of equation (\ref{stationarity}).
Estimates for $r$ and $p$ are also shown in table \ref{nwtab}.

\subsection{Characteristic exponents}

Simulation results for the values of the characteristic exponents $\gamma, q$, and $\beta$ in 
the model depend on the parameters $p$ and $r$, as shown in figure \ref{slopes}. 
We fix $N=10^3$ and $m=0$. Results are averaged over 500 realizations for each parameter pair $(p,r)$.
All three scaling exponents, equations (\ref{ScalingPi}-\ref{ScalingC}), can be explained by the model.

Model exponents for $\gamma$ fall in the range $0<\gamma<1$, depending on $p$ and $r$, figure \ref{slopes}(a).
$\gamma$ is close to one for high $p$ and high $r$. 
The preferential attachment associated with triadic closure is therefore sub-linear.
The dependence of the exponent $q$ on both $p$ and $r$ is shown in figure \ref{slopes}(b).
Note that for $q=1$ the $q$-exponential is equivalent to the exponential.
Values of $q$ above (below) one indicate that the distribution decays slower (faster) than the exponential.
For small $p$ and large $r$, $q$ is significantly larger than one and  degree distributions are fat-tailed.
For large $p$ the values of $q$ approach one, independent of $r$.
Values for $\beta$ are close to zero for $r=0$ or $p$ going to $0$.
$\beta$ approaches a plateau at $\beta = -1$ for high values of $p$ and $r$, see figure \ref{slopes}(c).

\begin{figure}[tbp]
 \begin{center}
 \includegraphics[width=85mm]{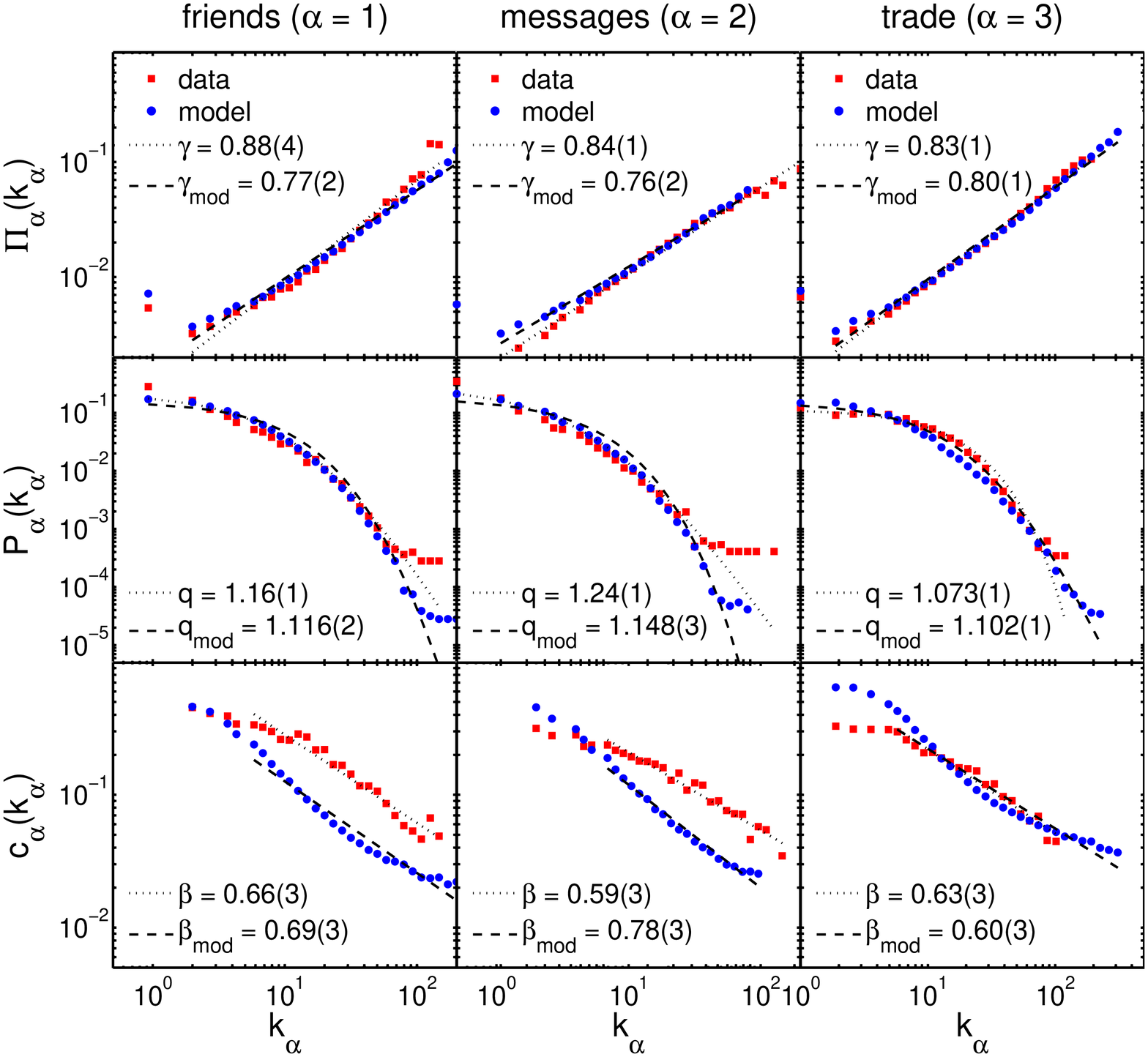}
  \end{center}
 \caption{Network scaling-exponents of the social multiplex can be explained by the calibrated model.
 Results are shown for the Pardus friendship ($\alpha=1$, left column), communication ($\alpha=2$, middle column), 
 and trade network ($\alpha=3$, right column). All data is logarithmically binned.
{\it Top row:} The attachment kernels scales sub-linearly with the node degrees in each case for data ($\gamma$) and model ($\gamma_{mod}$). 
Curves for data and model are barely distinguishable from each other.
{\it Middle row:} Degree distributions for $\alpha=1,2,3$ and best fits of a $q$-exponential, for data ($q$) and model ($q_{mod}$). 
{\it Bottom row} The scaling of the average clustering coefficients as a function of degree is compared between data and model.
Fits for $\beta$ and $\beta_{mod}$ yield almost the same results for friends and trades, with comparably larger deviations for the communication 
network. The model results for $c_{\alpha}(k_{\alpha})$ show an upwards curvature for high $k_{\alpha}$.}
  \label{data}
 \end{figure}

For the experimental validation of the model, figure \ref{data} shows the attachment kernel $\Pi_{\alpha}(k_{\alpha})$, 
degree distribution $P_{\alpha}(k_{\alpha})$, and clustering coefficients $c_{\alpha}(k_{\alpha})$ for the three sub-networks 
$M_{\alpha}$ of the empirical multiplex data.
They are compared to the respective distributions of the calibrated model (results  averaged over 20 realizations).
Data and model results are logarithmically binned, a version of figure \ref{data} showing raw data can be found in the supplementary information.

The observed preferential attachment in the data is in good agreement with  model results for each 
network $M_{\alpha}$, see top row of figure \ref{data}.
We find exponents of $\gamma=0.88(4)$ for the data and $\gamma_{mod}=0.77(2)$ in the model for the friendship network, 
$\gamma = 0.84(1)$, $\gamma_{mod} = 0.76(2)$ for communication, and $\gamma=0.83(1)$, $\gamma_{mod} = 0.80(1)$ for trade.
Data and model curves for $\Pi_{\alpha}(k_{\alpha})$ are barely distinguishable from each other.
The model fits the number of friends per player with exponents $q = 1.16(1)$ and $q_{mod} = 1.116(2)$ for 
$\alpha = 1$, $q = 1.24(1)$, $q_{mod} = 1.148(3)$ for $\alpha = 2$, and $q = 1.073(1)$ and $q_{mod} = 1.102(1)$ for $\alpha = 3$.
Results are shown in the middle row in figure \ref{data}.
Data and model show similar scaling of the average clustering coefficient of nodes $c_{\alpha}(k_{\alpha})$ as a function 
of  their degree $k_{\alpha}$, see bottom row in figure \ref{data}.
For friendships ($\alpha=1$) we find $\beta=0.66(3)$, for the model $\beta_{mod} = 0.69(3)$.
For communication ($\alpha=2$) the data yields $\beta = 0.59(3)$, the model gives $\beta_{mod} = 0.78(3)$.
For trade ($\alpha=3$) there is good agreement between data and model with $\beta = 0.63(3)$ and $\beta_{mod}=0.60(3)$, respectively.
The model results for $c_{\alpha}(k_{\alpha})$ show a curvature and are not straight lines.
Comparing the curves for $\alpha=1,2,3$ suggests that this curvature increases with the average degree $\bar k_{\alpha}$.
Values for $\beta_{mod}$ should be interpreted as first order approximations for the slopes of these curves.
Results for the exponents $\gamma, q, \beta$ for data and model are summarized in table \ref{nwtab}.

\section{\label{sec:discussion}Discussion}

We reported strong evidence that the process of triadic closure may play an even more 
fundamental role in social network formation than previously anticipated \cite{Rapoport53, Granovetter73}.
Given that {\em all} model parameters can be measured in the data, it is remarkable that three important 
scaling laws are simultaneously explained by this simple triadic closure model.
Since  exponents $\gamma$, $q$, and $\beta$ are sensitive to choices of the model parameters $p$ and $r$, 
the agreement between data and model is even more remarkable.

The Pardus multiplex data contains three other social networks, where links express negative relationships between players, 
such as enmity, attacks, and revenge \cite{Szell10}.
Triadic closure is known to be not a good network formation process for negative ties, 
"the enemy of my enemy is in general {\it not} my enemy" \cite{Heider46}.
It was shown that the probability of triadic closure between three players is one 
order of magnitude smaller for enmity links when compared to friendship links in the Pardus multiplex \cite{Szell10SN, Szell10}.
The model is therefore not suited to describe network formation processes of links expressing negative sentiments.


The findings in the current model also compare well to several facts of real-world social networks. 
Sub-linear preferential attachment has been reported in  scientific collaboration 
networks and the actor co-starring network ($\Pi(k) \propto k^{0.79}$ and $ \propto k^{0.81}$, respectively \cite{Jeong03}).
Degree distributions of many social networks often fall between exponential and power-law 
distributions \cite{Barabasi99,Newman01, Onnela07, Szell10SN, Amaral00}, and 
scaling of the average clustering coefficients as a function of  degree,  
has been observed in the scientific collaboration and actor networks with values for $c(k)\propto k^{-0.77}$ and $\propto k^{-0.31}$, 
respectively (when same fitting as in figure \ref{data} is applied). Mobile phone and communication 
networks give $\propto k^{-1}$ \cite{Onnela07a}.

In the Pardus dataset players are removed if they choose to leave the game or if they are 
inactive for some time \cite{Szell10SN}. In the mobile communication, 
actor, and collaboration networks, a link is established by a single action 
(phone call, movie, or publication) and persists from then on.
Note that our model addresses the empirically relevant case where node-turnover rates 
($\Delta n^+_{\alpha}, \Delta n^-_{\alpha}$) are significantly larger than the effective 
network growth rate ($\Delta n^+_{\alpha} - \Delta n^-_{\alpha}$).
For growing networks (without node deletion) it has been shown that sub-linear 
preferential attachment ($\gamma < 1$) leads to degree distributions 
with power-law tail with an exponent proportional to $\gamma$ \cite{Krapivsky00}.
Something similar can be observed in the present model.
If we keep the node-turnover parameter $p$ fixed and decrease the triadic closure parameter $r$, 
figures \ref{data}(a) and (b) show that $\gamma$ decreases and $q$ approaches one.
The network is dominated by randomly created links.
However, if we fix $r=1$ (only triadic closure, no random links) and increase $p$,
figures \ref{data}(a) and (b) show that $q$ approaches one despite an increase in $\gamma$.
An increase of the node-turnover parameter $p$ implies a shorter life-time for individual nodes and hence a shorter time in which they may acquire new links.
Consequently, the degree distribution only has a substantial right-skew if both, $p \lesssim 0.25$, and $r \gtrsim 0.5$ holds.

\section{\label{sec:methods}Methods}

\subsection{Multiplex data}

The Pardus dataset allows to continuously track all actions of more than 370,000 players in an open-ended, 
virtual, futuristic game universe where players interact  in a multitude of ways to achieve their self-posed goals, 
such as accumulating wealth and influence. Players can establish friendship links, exchange one-to-one messages 
(similar to phone calls) and trade with each other.
We focus on three sub-networks (friendship, communication, trade) of the multiplex, over one year from Sep 2007 to Sep 2008.
Network label $\alpha=1$ refers to the friendship network, $\alpha=2$ for communication, and $\alpha=3$ for trade.
In the friendship network a node is present on a given day if at least one friendship link to another node exists on that day.
A node is removed if the player either leaves the game or has no friendship link.
The same holds for the message and trade networks, where a link exists between two nodes on day $t$ 
if at least one message (trade) is exchanged within the period of six days, $\left[ t-6,t \right]$.
For details of structural and dynamical properties of the Pardus multiplex, see \cite{Szell10, Szell10SN, SzellChapter, Thurner12, Szell12}.

To measure the degree distributions $P_{\alpha}(k_{\alpha})$ and clustering coefficients $c_{\alpha}(k_{\alpha})$, 
we use the adjacency matrix of the networks $M_{\alpha}$ on the last day of the data record.
The preferential attachment probability $\Pi_{\alpha}(k_{\alpha})$ is measured by counting (over the entire observation period)
the number of link-creation events in which a node with degree $k$ acquires a new link, and then dividing this by the average 
number of nodes with degree $k$, where the average is again taken over the observation period. 

\subsection{Fitting procedures}

Power-law fits (least-squares) to the logarithms of the logarithmically binned data in figure \ref{data} are shown for ${\gamma}$, for $2<k_{(\alpha)}<100$, and for ${\beta}$ over the range 
$5<k_{(\alpha)}<100$, for each ${\alpha}$, for data and model. The reported errors are the standard deviations of the coefficients.
For the degree distributions the data is also logarithmically binned and fitted over the entire range $k_{(\alpha)}>0$ in figure \ref{data} with equation (\ref{ScalingK}).
The coefficients are obtained as maximum likelihood estimates, reported errors correspond to the 95\% confidence intervals.
For better comparison and to diminish the effect of outliers, data and model results for $\Pi_{\alpha}(k_{\alpha})$ 
are normalized over the range $k_{\alpha} \leq 100$.
Higher values correspond to data outliers, often due to behavior of non-serious players. 

\section{Acknowledgments}
This work was supported by Austrian science fund FWF P23378 and EU FP7 projects CRISIS no. 288501 and LASAGNE no. 318132. 
We thank  B. Fuchs and M. Szell for data issues.

\pagebreak

\section{Supplementary Information}

 \begin{figure}[bp]
 \begin{center}
 \includegraphics[width=120mm]{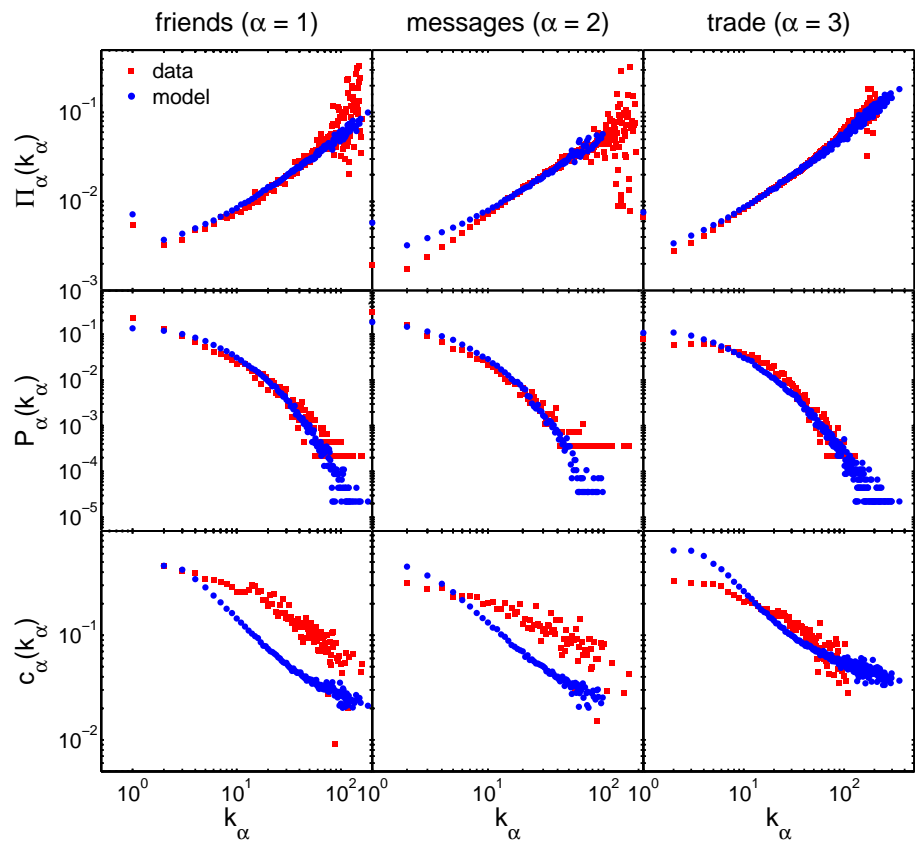}
  \end{center}
 \caption{Comparison between data and model results for the Pardus friendship ($\alpha=1$, left column), communication ($\alpha=2$, middle column), 
 and trade network ($\alpha=3$, right column).
{\it Top row:} The attachment kernels scales sub-linearly with the node degrees in each case for data and model. 
Curves for data and model are barely distinguishable from each other.
{\it Middle row:} Degree distributions for $\alpha=1,2,3$ for data and model. 
{\it Bottom row} The scaling of the average clustering coefficients as a function of degree is compared between data and model.}
  \label{data2}
 \end{figure}

\end{document}